\documentclass{iaus}
\usepackage{graphicx}

\def\be{\begin{equation}}
\def\ee{\end{equation}}
\def\bea{\begin{eqnarray}}
\def\eea{\end{eqnarray}}

\def\lapprox{\mathrel{\hbox{\rlap{\hbox{\lower4pt\hbox{$\sim$}}}\hbox{$<$}}}}
\def\gapprox{\mathrel{\hbox{\rlap{\hbox{\lower3pt\hbox{$\sim$}}}\hbox{$>$}}}}

\title[Coherent emission] 
{Coherent emission}

\author[D.B. Melrose]   
{D.B. Melrose}
\affiliation{School of Physics, University of Sydney,\\ NSW 2006, Australia \\email: {\tt melrose@physics.usyd.edu.au}}

\pubyear{2008}
\volume{257}
\pagerange{TBA}
\jname{Universal Heliophysical Processes}
\editors{N. Gopalswamy \&  D.F. Webb eds.}
\begin{document}

\maketitle

\begin{abstract}
The theory of plasma emission and of electron cyclotron maser emission, and their applications to solar radio bursts and to Jupiter's DAM and the Earth's AKR are reviewed, emphasizing the early literature and problems that remain unresolved. It is pointed out that there are quantitative measures of coherence in radio astronomy that have yet to be explored either observationally or theoretically.
\keywords{radiation mechanisms: nonthermal, Sun: radio radiation, interplanetary medium.}
\end{abstract}

\firstsection 
\section{Introduction}

It is 50 years this year since the first papers were published on two radio emission processes that have become central to our understanding of radio emission in the heliosphere: plasma emission \cite[(Ginzburg \& Zheleznyakov 1958)]{GZ58}, and electron cyclotron maser emission\cite[(Twiss 1958)]{T58}. Ginzburg \& Zheleznyakov's work provided a theoretical framework for a qualitatively accepted idea: solar radio bursts (types~I, II, III were then known) are due to `plasma emission' in which Langmuir waves are excited and secondary processes produce escaping radiation near the plasma frequency and its second harmonic. Plasma emission is now well established for other heliospheric radio emissions, notably from shocks and from planetary magnetospheres. In contrast, Twiss' theory (which he applied to type~I bursts) was largely ignored; it was about two decades later that electron cyclotron maser emission (ECME) became accepted as the emission process for Jupiter's DAM, the Earth's AKR and related emissions from the outer planets. 

Plasma emission and ECME are two examples of what is called `coherent' emission. (There is a third astrophysical coherent emission process, pulsar radio emission, which is inadequately understood and is not discussed here.) The characteristic feature used to distinguish coherent from incoherent emission is the high brightness temperature, $T_B$: any emission that is too intrinsically bright to be explained in terms of an incoherent emission process is assumed to involve a coherent emission process. There are three classes of coherent emission processes (\cite[Ginzburg \& Zheleznyakov 1975]{GZ75}; \cite[Melrose 1986a]{M86a}): antenna mechanisms, reactive instabilities and maser instabilities. In an antenna mechanism it is assumed that there exists a bunch of $N$ electrons, localized in both coordinate space, ${\bf x}$, and in momentum space, ${\bf p}$, such that the bunch radiates like a macro-charge, leading to a power $N^2$ times the power that an individual electron would radiate. The back-reaction to the emission increases the spread in ${\bf x}$, leading to self-suppression. In a reactive instability, there is assumed to be localization in ${\bf p}$, but not in ${\bf x}$: wave growth is due to a feedback mechanism that leads to self-bunching, causing a phase-coherent wave to grow. The growth tends to increase the spread in ${\bf p}$, leading to self-suppression when this spread causes the bandwidth to exceed the growth rate. A maser instability involves an inverted energy population, with growth corresponding to negative absorption; the back reaction, described quantitatively in terms of quasilinear relaxation, reduces the inverted population, leading to self-suppression.

Which of these coherent emission processes is appropriate in astrophysical and space plasmas?  In space plasmas the time available for wave growth is typically very much longer than the growth times and one expects the instabilities to saturate quickly. An antenna mechanism should saturate and evolve into a reactive instability, a reactive instability should saturate and evolve into a maser instability, which should saturate and reach a state of marginal stability. Rapid growth and saturation is necessarily confined to a short time and a small volume, and is expected to produce only highly localized transient bursts of coherent emission. For coherent emission to be observable in radio astronomy, it must come from a sufficiently large volume over a sufficiently long time. In the absence of evidence to the contrary, this argument suggests that only maser instabilities are likely to be relevant and one expects the electron distribution to relax quickly to a marginally stable configuration, minimizing the growth rate of the instability.

After reviews of plasma emission (\S\ref{sect:plasma}) and ECME (\S\ref{sect:ECME}), I discuss two aspects of coherent emission that require further development (\S\ref{sect:coherence}).

\section{Plasma emission}
\label{sect:plasma}

\begin{figure}[t]
\begin{center}
 \includegraphics[width=3.4in]{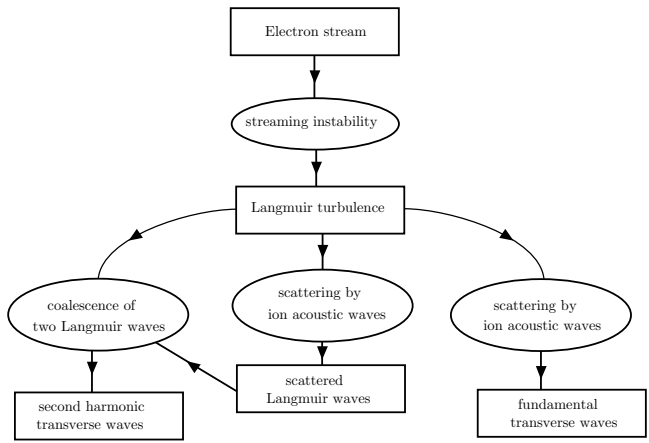} 
 \caption{A flow diagram indicating the stages in plasma emission in an updated version on the original theory (\cite[Melrose 1970a\&b]{M70a,b}:  \cite[Zheleznyakov \& Zaitsev 1970a\&b]{ZZ70}).}
   \label{fig1}
\end{center}
\end{figure}

The essential idea in plasma emission (notably in type~III bursts) is that an electron stream (electron `beam') generates Langmuir (L) waves, and that these lead to emission of escaping radiation at the fundamental (F) and second harmonic (H) of the plasma frequency, as indicated in the flow diagram Fig.~\ref{fig1}. The remarkable contribution of \cite[Ginzburg \& Zheleznyakov (1958)]{GZ58} was that they developed their theory before much of the relevant plasma theory had been properly developed for laboratory plasmas.

\subsection{Theory of plasma emission}

The first stage in plasma emission involves the beam instability that produces the L-waves. Ginzburg \& Zheleznyakov appealed to a reactive version of the beam instability, and this stage was later updated to the slower-growing maser version, sometimes called the bump-in-tail instability. Even this instability grows extremely rapidly compared with the time scale of type~III bursts, leading to a dilemma, pointed out by \cite[Sturrock (1964)]{S64}. During quasilinear relaxation of the beam, it loses a substantial fraction of its energy, implying that the beam should stop, over a distance that Sturrock estimated in kilometers. This is clearly not the case: type~III streams can propagate through the corona, and through the  interplanetary medium (IPM) to several AU. One suggested way of overcoming this dilemma involved a recycling in which the faster electrons at the front of the beam lose energy to the L-waves, and the slower electrons at the back of the beam reabsorb them. This effect does indeed seem to occur (\cite[Grognard 1983]{GR83}), but the energy loss remains catastrophic unless the efficiency of the recycling is unrealistically close to 100\%, whereas one expects the Langmuir waves to evolve due nonlinear (strong-turbulence) processes (\cite[Goldman 1983]{G83}) before they can be reabsorbed. The actual resolution of Sturrock's dilemma emerged from {\it in situ\/} observations of type~III burst in the IPM near 1AU.

The first spacecraft to observed type~III bursts {\it in situ\/} in the IPM initially found no L-waves (\cite[Gurnett \& Frank 1975]{GF75}; \cite[Gurnett \& Anderson 1977]{GA77}), trivially resolving the dilemma, but seemingly undermining the theory of plasma emission. Subsequently, it was recognized that L-waves are present, but confined to isolated clumps, implying that the growth of L-waves occurs only in highly localized, transient bursts. The favored explanation is that the beam is in a state of marginal stability. The tendency to growth increases systematically due to faster electrons overtaking slower electrons, tending to increase the positive gradient in momentum space near the front of the beam. This effect is offset by quasilinear relaxation, in the many isolated clumps of L-waves, tending to reduce this gradient. Balancing these two effects leads to marginal stability. Qualitatively, in the marginally stable state, various weak damping processes mostly prevent effective growth, which occurs only in isolated regions where the conditions are particularly favorable for growth. This suggests a `stochastic growth' model in which many such isolated, localized bursts of growth occur. A specific stochastic growth model, in which the amplification factor in each burst is $\exp G$ with $G$ obeying gaussian statistics, predicts a log-normal statistical distribution of electric fields, and this accounts well for the observed statistical distribution \cite[(Robinson, Cairns \& Gurnett 1993)]{RCG93}.

It is now recognized that there is a rich variety of specific processes that can partially convert the energy in Langmuir waves into escaping F and H radiation. In the flow diagram Fig.~\ref{fig1}, it is assumed that low-frequency waves, such as ion sound (S) waves, are also excited, and these play two roles. Coalescence and decay processes involving L- and S-waves can produce F emission, $\rm L\pm S\to F$, and backscattered L-waves, $\rm L\pm S\to L'$, which coalesce with the beam-generated waves to produce H emission, $\rm L+L'\to H$. 

\subsection{Application to solar radio bursts}

Several problems arose in the early interpretation of the escape of solar radio emission from the corona:
\begin{itemize}
\item refraction for F emission implies that only sources within about a degree of the central meridian should be visible;
\item modeling of the propagation of type~III burst produced consistent results only for the standard coronal density models multiplied by a large factor, typically 30 or 100;
\item the radio temperature of the quiet Sun is $\sim1/10$ of the known temperature;
\item F and H emission at a given frequency appear to come from a given height;
\item sources at a given frequency have a similar apparent size.
\end{itemize}
A seemingly plausible interpretation is that scattering off coronal inhomogeneities increases the  cone-angle of the escaping radiation and results in a scatter-image with an apparent area much larger than the actual area of the source. However, this violates a fundamental constraint (a Poincar\'e invariant, or generalized \'etendue) that the product of the apparent area and the cone angle is a constant for rays propagating through any optical system. This dilemma is resolved by a model in which the radiation is ducted outward in the corona (\cite[Duncan 1979]{D79}) by reflecting off highly collimated overdense structures, called fibers by \cite[Bougeret  \& Steinberg (1977)]{BS77}. It is not widely appreciated that the resolution of these problems in the interpretation of the radio data require that the corona be extremely inhomogeneous and highly structured on scales much smaller than can be resolved by present techniques.

Although type~III bursts are relatively well understood, there are unresolved problems in identifying the exciting agencies and specific properties of other types (e.g., \cite[McLean \& Labrum 1985]{ML85}). Type~II bursts are perhaps the next best understood, being the most type~III-like, but how the shock wave produces type~III-like electron streams is still uncertain (\cite[Knock \& Cairns 2005]{KC05}). Type~I emission is less understood; type~I emission can include both bursts and a continuum, neither of which are related to flares. No definitive model for type~I emission has emerged, despite an enormous amount of detailed data (\cite[Elgar\o y 1977]{E77}). It is unclear what the exciting agency is for the type~I continuum and indeed for any other broad-band plasma emission such as the flare continuum (e.g., \cite[McLean \& Labrum 1985]{ML85}). 

Observationally, plasma emission is circularly polarized, and various problems arose in connection with the interpretation of the polarization. Variants of the flow diagram Fig.~\ref{fig1}, involve including the effect of the magnetic field on the low-frequency waves (replacing the S-waves) and on the F and H emission, which are in the magnetoionic modes. Circular polarization is interpreted as an excess of one magnetoinic wave over the other, and observationally plasma emission favors the o~mode over the x~mode. An important qualitative result concerning the polarization is that conventional F emission processes produce radiation between the plasma frequency, $\omega_p$, and the cutoff frequency of the x~mode, $\omega_x=\Omega_e/2+(\omega_p^2+\Omega_e^2/4)^{1/2}$. In this frequency range only the o~mode can propagate, and hence F emission should be 100\% polarized in the o~mode. This is the case for most type~I emission, but F emission in type~III and type~II bursts, although polarized in the o~mode, is never 100\%. Also the polarization of type~I emission decreases as the source approaches the solar limb (\cite[Zlobec 1975]{Z75}). This is interpreted as a depolarization due to propagation effects in the inhomogeneous corona. Reflection off duct walls can lead to such depolarization, but only if there are extremely sharp density gradients at the edges of the fibers (\cite[Melrose 2006]{M06}).

\section{ECME}
\label{sect:ECME}

The favored interpretation of Jupiter's decametric radio emission (DAM) and the Earth's auroral kilometric radiation (AKR) is in terms of loss-cone driven ECME.

\begin{figure}[t]
\begin{center}
 \includegraphics[width=2.5in]{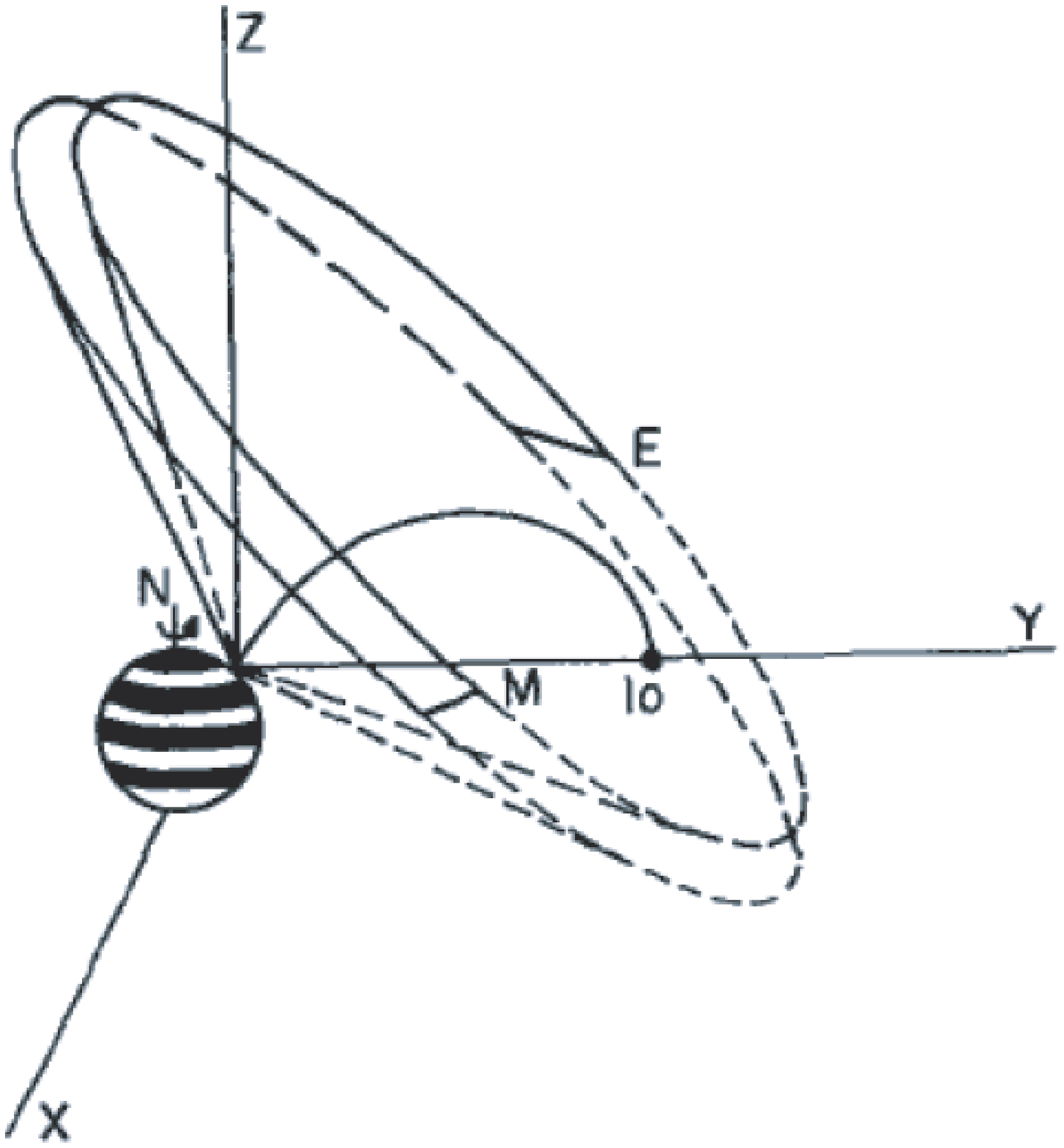} 
 \caption{The angular pattern for DAM (\cite[Dulk 1967]{D67}).}
   \label{fig:dulk67}
\end{center}
\end{figure}

\subsection{DAM and AKR}

Early interest in DAM (\cite[Burke \& Franklin 1955]{BF55}) increased dramatically following a report  \cite[(Bigg 1964)]{B64} that the radio bursts correlate with the position of the innermost Galilean satellite, Io. An explanation for the Io effect was formulated a few years later (\cite[Piddington \& Drake 1968]{PD68}), and described as a unipolar inductor by \cite[Goldreich \& Lynden-Bell (1969)]{GL69}. The idea is that the Jovian magnetic field lines that thread Io are frozen into Io and dragged through the corotating Jovian magnetosphere. The relative motion of magnetic field lines sets up an EMF of a few MV that drives a field-aligned current, with the circuit closing across Io and in the Jovian ionosphere. The EMF accelerates electrons up to a few MeV somewhere along the Io flux tube. A major component of DAM is emitted just above the Jovian ionosphere on the Io flux tub.  The angular emission pattern of this radiation seems bizarre: confined to the surface of a hollow cone with its apex on the Io flux tube, with an opening angle of the cone $\approx 80^\circ$ and a thickness $\approx 1^\circ$, cf. Fig.~\ref{fig:dulk67}. DAM is attributed to cyclotron emission by accelerated electrons that mirror and propagate upward with a loss-cone anisotropy. The fly-bys of Jupiter by the Pioneer and Voyager spacecraft largely confirmed the cyclotron interpretation of DAM, notably a pattern of nested arcs about the north Jovian magnetic pole gave strong support to the model of emission on the surface of a hollow cone.

\begin{figure}[t]
\begin{center}
 \includegraphics[width=2.5in]{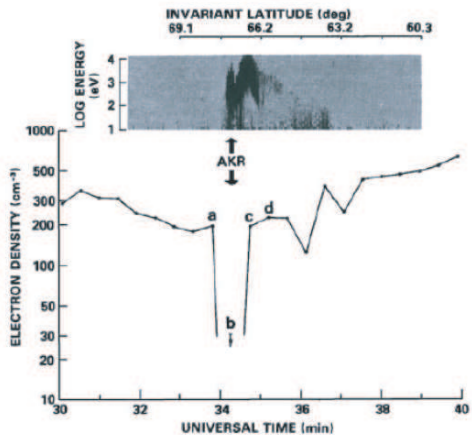} 
\caption{The electron density drops dramatically (to $\omega_p/\Omega_e<0.2$) as a function of time as the spacecraft (Isis 1) crosses the source region of AKR, indicated by the insert showing the dynamic spectrum of inverted-V electrons \cite[(Benson \& Calvert 1979)]{BC79}. }
\label{fig:cavity}
\end{center}
\end{figure}

The discovery of AKR in the early 1970s (\cite[Gurnett 1974]{G74}) implied that the Earth is a spectacular radio source below the ionospheric cutoff frequency. The source region for AKR is above the auroral zones and near the last closed field line. The polarization is dominantly x~mode. AKR correlates with `inverted-V' electron precipitation events, with energies $\approx 2$ to $10\rm\, keV$. Data from spacecraft that pass directly through the source show that these electrons are accelerated by a parallel potential gradient, and that the inverted-V events occur in density cavities, cf.\ Fig.~\ref{fig:cavity}. The cavities are localized to regions of strong upward current, that draws electrons downward and exhausts the magnetospheric supply of thermal electrons on the current-carrying flux tube. This leads to `charge starving', so that the the available EMF localizes and accelerates all the available electrons to several keV.

To explain AKR and DAM one needs to account for the following features.\\
(1) The emission is near the local cyclotron frequency, with a very high $T_B$. \\
(2) The emission is predominantly in the x mode of magnetoionic theory.\\
(3) The emission pattern for DAM is confined to a thin
surface of a wide hollow cone.\\
(4) AKR correlates with inverted-V precipitating electrons.

\subsection{Theory of ECME}

Electron-cyclotron instability can be classified into four types (\cite[Melrose 1986a]{M86a}): $\perp$-driven and $\parallel$-driven maser instabilities (\cite[Twiss 1958]{T58}; \cite[Schneider 1959]{S59}), and reactive instabilities due to azimuthal and axial self-bunching (\cite[Zhelznyakov 1959]{Z59}; \cite[Gaponov 1959]{G59}).  The favored mechanism for ECME in astrophysics is a perpendicular-driven maser, which (like the azimuthal bunching instability) depends on an intrinsically relativistic effect, even for electrons otherwise regarded as nonrelativistic.

The intrinsically relativistic effect in $\perp$-driven ECME can be understood by considering the gyroresonance condition
\be
\omega-s\Omega_e/\gamma-k_\parallelÊv_\parallel= 0,	
\label{ecme1}
\ee
where $sÊ=0,\pm1,\pm2,\ldots$ is the harmonic number, $\Omega_e=eB/m$ is the electron cyclotron frequency, and $\gamma$ is the Lorentz factor. Suppose one considers all the electrons that can resonate at given $s$, frequency $\omega$, and parallel wavenumber $k_\parallel$. Let the electrons be described by their perpendicular and parallel velocity components, $v_\perp$, $v_\parallel$. The resonant electrons lie on a resonance ellipse (\cite[Omidi \& Gurnett 1982]{OG82}; \cite[Melrose, R\"onnmark \& Hewitt 1982]{MRH82}) in $v_\perp$-$v_\parallel$ space. The absorption coefficient involves an integral around the resonance ellipse, and the sign of the integrand is negative for
\be
\left[{s\Omega_e \over v_\perp}{\partial\over \partial p_\perp}
+k_\parallel{\partial\over \partial p_\parallel}\right]
f(p_\perp,p_\parallel) > 0 .		
\label{driver}
\ee
For a given unstable distribution of electrons, the maximum growth rate corresponds to the specific resonance ellipse that gives the maximum positive value of this integral.

Suppose one makes the nonrelativistic approximation, $\gamma=1$, in (\ref{ecme1}); then the resonance ellipse reduces to a line at $v_\parallel=(\omega - sÊ\Omega_e)Ê/k_\parallel$. In this case, the integral along the line $p_\parallel=$ constant may be partially integrated, and one can show that the net contribution to the absorption coefficient from the $p_\perp$-derivative cannot be negative. In this approximation only $\parallel$-driven maser action is possible (\cite[Melrose 1976]{M76}). However, the requirements on the distribution function for effective growth are rather extreme, and not satisfied for the measured inverted-V distribution function.

An emphasized by \cite[Twiss (1958)]{T58}, the $\perp$-driven maser involves an intrinsically relativistic effect, even for electrons that one would not regard as relativistic (e.g., a few keV).  In the case of perpendicular propagation, $k_\parallel\to0$,  (\ref{ecme1}) with $1/\gamma\approx1+v^2/2c^2$ implies that the resonance ellipse becomes a circle centered on the origin. The integral of the $p_\perp$-derivative around this circle can be negative, for example, for a `shell' distribution with $\partial f/\partial p>0$, which was the case considered by Twiss. For nearly perpendicular propagation the resonance ellipse is approximately circular, with a center displaced from the origin along the $v_\parallel$-axis by a distance $\propto k_\parallel$. The relativistic effect cannot be ignored when the radius of the resonance circle is comparable with the speed of the electrons that drive the maser. The paradox that one must include the relativistic correction to treat $\perp$-driven ECME, even for nonrelativistic electrons, is resolved by noting that the nonrelativistic approximation is formally $c\to\infty$, whereas $c$ is necessarily finite when treating electromagnetic radiation.

\begin{figure}[t]
\begin{center}
 \includegraphics[width=2.5in]{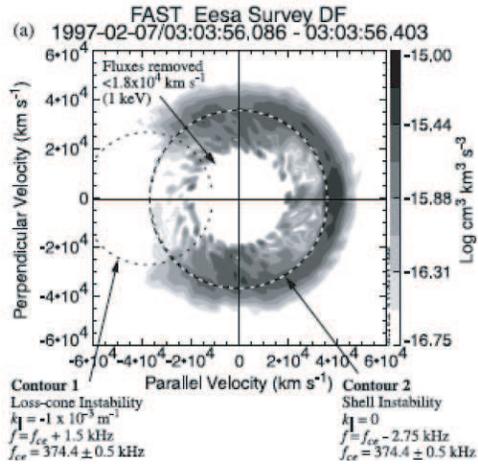} 
\caption{The electron distribution in an inverted-V event; two resonance ellipses that correspond to negative absorption are shown, one inside the loss cone on the left and the other centered on the origin passing through regions with $\partial f/\partial p>0$ associated with the `horseshoe' or `shell' distributions (\cite[Ergun et al. 2000]{Eetal00}).}
\label{fig:ecme4}
\end{center}
\end{figure}

\cite[Wu \& Lee (1979)]{WL79} proposed a loss-cone driven instability for AKR. The maximum growth rate is for the resonance ellipse that lies just inside the loss cone, and samples the region where $\partial f/\partial p_\perp$ has its maximum positive value, as illustrated in Fig.~\ref{fig:ecme4}. A major success for this theory was that the inverted-V electrons were found to have a loss-cone distribution of the required form.  Loss-cone driven ECME leads naturally to narrowband, fundamental x-mode ECME restricted to a narrow surface of a wide hollow cone. This may be understood by noting that the position of the center of the resonance ellipse determines the angle of emission, and that the growth rate is a sensitive function of the position of the center of the ellipse.  This property explains the seeming bizarre angular distribution of DAM in Fig.~\ref{fig:dulk67}.  The frequency is determined by the semi-major axis of the ellipse, and a similar argument implies that growth is confined to a narrow bandwidth, centered on the Doppler-shifted cyclotron frequency.

Besides the loss-cone feature, there are other features in the measured distribution function of inverted-V electrons that could produce ECME, specifically a trapped distribution (\cite[Louarn et al. 1990]{Letal90}) and shell and horseshoe features  (\cite[Bingham \& Cairns 2000]{BC00}; \cite[Ergun et al. 2000]{Eetal00}). As in Twiss' theory, growth in these cases occurs at $\Omega_e/\gamma<\Omega_e$, and in a cold plasma such radiation is in the z~mode, below a stop band between $\Omega_e$ and $\omega_x$ that precludes escape. The recognition that there is essentially no cold plasma inside the cavity overcomes this difficulty, with warm plasma effects washing out the stop band; the cavity acts like a duct, with AKR reflecting off the sides until it reached a height where $\omega_x$ outside the duct is low enough to allow escape (\cite[Ergun et al. 2000]{Eetal00}). Due to the absence of cold plasma, trapped, shell or horseshoe distributions are viable alternatives to a loss-cone distribution as the driver of ECME for AKR. However, they cannot account for the angular distribution of DAM, Fig.~\ref{fig:dulk67}.

\subsection{Stellar applications of ECME}

Loss-cone driven ECME is a candidate for emission in any other astrophysical context where two conditions are satisfied: energetic electrons precipitate in a magnetic bottle, and the ratio $\omega_p/\Omega_e$ is sufficiently small. One such context is in solar microwave `spike' bursts (\cite[Holman, Eichler \& Kundu1980]{HEK}; \cite[Melrose \& Dulk 1982]{MD82}).  These bursts turn on and off in around $10\rm\, ms$, and have high brightness temperatures ($\approx10^{15}Ê\rm\,ÊK$).  Another application is to the interpretation of radiation from some radio flare stars.  There are several classes of flare stars, including M-type (red dwarf) stars, also called UV Ceti variables, which have extremely powerful solar-like flares,  and cataclysmic variable stars, which involve mass transfer onto a magnetized star from a binary companion.

A major unsolved problem is how ECME escapes from a stellar corona. Along any prospective escape path, x-mode radiation generated just above $\omega_x$ encounters a layer where the local value of $\Omega_e$ is equal to $\omega/2$, where second-harmonic cyclotron absorption by thermal electrons is very strong.  Four suggested ways of overcoming or avoiding the strong absorption at $2\Omega_e$ have been suggested.\\
(1) ECME at $s\ge2$, or in the o~mode (\cite[Melrose, Hewitt \& Dulk 1984]{MHD84}; \cite[Winglee 1985]{W85}). \\
(2) When fundamental x~mode emission is not allowed, ECME favors fundamental z~mode emission, and the escaping radiation results from coalescence of z~mode waves to produce higher harmonic x~mode emission (\cite[Melrose 1991]{M91}). \\
(3) There are `windows' at parallel or perpendicular propagation where some radiation can escape  (\cite[Robinson 1989]{R89}).\\
(4) The radiation can tunnel through the layer: the radiation incident on the absorbing layer from below, with $\omega<2\Omega_e$, modifies the distribution function of the electrons, allowing re-emission at $\omega>2\Omega_e$ (\cite[McKean,  Winglee \& Dulk 1989]{MWD89}). \\
One possibility is that ECME generates a large power, with only a tiny fraction of it escaping. An interesting implication is that the absorption of this large power can lead to non-local heating at the absorption layer. These suggested stellar applications of ECME will remain questionable until the escape of the radiation is explained convincingly.

\section{Two outstanding problems}
\label{sect:coherence}

Our understanding of coherent emission is far from complete. I comment on two aspects: phase-coherent growth and measures of coherence.

\begin{figure}[t]
\begin{center}
  \includegraphics[width=5.0in]{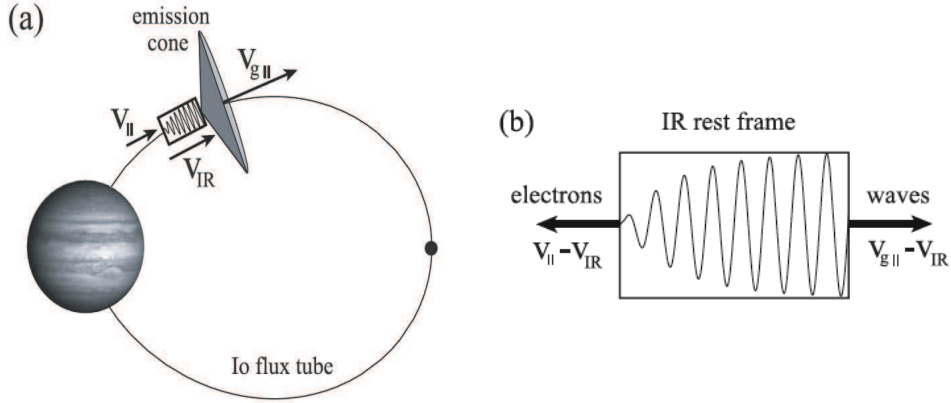} 
\caption{The phase-coherent model for ECME proposed to explain fine structure in DAM: the interaction region is moving at $V_{\rm IR}$, the wave amplitude is illustrated inside the box, with the waves propagating at the group velocity $V_{\rm g\parallel}$, and the resonant electrons at $V_\parallel$ (\cite[Willes 2002]{W02}).}
\label{ecme:fig4}
\end{center}
\end{figure}

\subsection{Phase-coherent wave growth}

Maser growth applies only if the growth rate is less than the bandwidth of the growing waves. There are examples where fine structures are observed apparently inconsistent with this condition: fine structure in solar radio bursts \cite[(Ellis 1969)]{E69}, in S-bursts in DAM (\cite[Ellis 1973]{E73}; \cite[Carr \& Reyes 1999]{CR99}). In these examples the frequency tends to change with time, somewhat analogous to the narrow-band drifting structures in triggered VLF emissions \cite[(Helliwell 1967)]{H67}. It is unclear how such fine structures are related to broader band emission that is consistent with maser theory. The most extreme example of fine structures is in giant pulses the Crab pulsar (\cite[Hankins \& Eilek 2007]{HE07}); whereas pulsar emission typically satisfies log-normal statistics, giant pulses do not (\cite[Cairns, Johnston \& Das 2004]{CJD}). In all three cases, it seems that fine structures involve some distinctive physical process different from other coherent emission. One suggestion, illustrated in Fig.~\ref{ecme:fig4}, is a modified form of Helliwell's \cite[(1967)]{H67} phenomenological model for VLF emissions. The idea is that a reactive instability grows relatively slowly within an interaction region that is moving along the magnetic field lines, Fig.~\ref{ecme:fig4}.  On setting $\Delta=\omega-s\Omega_e/\gamma-k_\parallelÊv_\parallel$ one requires not only $\Delta=0$, which is the condition (\ref{ecme1}), but also $d\Delta/dt = 0$ along the path of the interaction region to determine how $\omega$ changes with $t$.

\subsection{Measures of coherence}

Is there some systematic way of quantifying the concept of coherence in radio astronomy? The only widely accepted measures of coherence in radio astronomy are either based on a high $T_B$ or on special techniques of high time- and frequency-resolution. However, there are other possible measures that are ignored. Historically, it is interesting that these other measures were initially transferred from radio physics to optics, notably through the Hanbury Brown-Twiss effect (e.g.,\cite[Hanbury Brown \& Twiss 1956]{HBR56}), and are now familiar in optics, in terms of photon-counting statistics, but unfamiliar in radio astronomy.

The measures of coherence involve powers of the intensity (e.g., \cite[Mandel \& Wolf 1995]{MW95}) averaged over a short observation time. The mean value of $\langle I^N\rangle$ for
``coherent''  radiation is $\langle I^N\rangle= \langle I\rangle^N$, and for ``incoherent'' (random phase) radiation is $\langle I^N\rangle= N!\langle I\rangle^N$. The quantities $\langle I^N\rangle/\langle I\rangle^N$ are measures of coherence. These measures must depend on the resolution time of the telescope and the coherence time of the radiation, and one expects interesting results only if the resolution time is shorter than the coherence time (which is not known {\it a priori}). There are additional measures of coherence from the auto- and cross-correlation functions between the Stokes parameters for polarized radiation. (In fact, the stated result for $\langle I^N\rangle$ is for completely polarized radiation, and there is a different value for partially polarized or unpolarized radiation.) At present there is no program for measuring these quantities, for example, by constructing a correlator that automatically determines these higher order correlation functions. There is also no systematic theory predicting what one would expect for various models of coherence.

\section{Conclusions}
\label{sect:conclusions}

Following the first papers in the field 50 years ago, major progress was made through the 1970s in the theory of plasma emission and its application to solar radio emission, and in ECME and its application to DAM and AKR about a decade later. There is  a renewed observational interest in meter-$\lambda$ radio astronomy (e.g., through LOFAR and MWA) and it is timely to re-appraise some of the unresolved problems left over from this earlier period. Some of these problems are outlined briefly above. There are also general problems relating to coherent emission {\it per se}. There are examples of fine structures in coherent emission that are inconsistent with maser theory, and appear to involve some intrinsically phase-coherent emission mechanism. 

It is pointed out that, in principle, there are measures of coherence available from correlation functions of powers of the intensity and the other Stokes parameters. It is possible to measure these quantities provided that the resolution time of the observations is shorter than the coherence time of the radiation.

\acknowledgement
I thank Matthew Verdon for comments on the manuscript.

\end{document}